\documentclass[12pt, draftclsnofoot, twocolumn]{IEEEtran}

\usepackage{cite}
\usepackage[utf8]{inputenc}
\usepackage{amsfonts}
\usepackage{amsmath}
\usepackage{mathrsfs}
\usepackage{graphicx}
\usepackage{booktabs}
\usepackage{makecell}
\usepackage{textcomp}
\usepackage{balance}
\usepackage[table,xcdraw]{xcolor}
\usepackage[a4paper,
            top=23mm,
            bottom=23mm,
            left=23mm,
            right=23mm]{geometry}

\newcolumntype{C}[1]{>{\centering\let\newline\\\arraybackslash\hspace{0pt}}m{#1}}
\usepackage{multirow}
\usepackage{soul,color}
\usepackage{array}

\def\BibTeX{{\rm B\kern-.05em{\sc i\kern-.025em b}\kern-.08em
    T\kern-.1667em\lower.7ex\hbox{E}\kern-.125emX}}
    
\usepackage{hyperref}
\hypersetup{
    colorlinks=true,
    linkcolor=magenta,
    filecolor=blue,      
    urlcolor=cyan,
    citecolor=green,
}

\begin{document}

\title{A Multi-Stage Hybrid CNN-Transformer Network for Automated Pediatric Lung Sound Classification}

\author{Samiul~Based~Shuvo\textsuperscript{1}, 
        Taufiq~Hasan\textsuperscript{1,*},~\IEEEmembership{Member,~IEEE}%
\thanks{\textsuperscript{1} Department of Biomedical Engineering, BUET, Dhaka, Bangladesh. 
Email: sbshuvo@bme.buet.ac.bd, taufiq@bme.buet.ac.bd}%
\thanks{\textsuperscript{*}Corresponding author}}

\markboth{XX, VOL. XX, NO. XX, XXXX 20XX}%
{Shuvo \MakeLowercase{\textit{et al.}}: Preprint}

\maketitle

\begin{abstract}

\noindent \textbf{Background:} Automated analysis of lung sound auscultation is essential for monitoring respiratory health, particularly in regions with a shortage of skilled healthcare workers. Although respiratory sound classification has been widely studied in adults, its application in pediatric populations, especially in children under six years of age remains underexplored. Developmental changes in pediatric lungs substantially modify the acoustic properties of respiratory sounds, requiring classification approaches tailored specifically to this age group. \\
\textbf{Methods:} To address this challenge, we propose a multistage hybrid CNN–Transformer framework that integrates CNN-extracted features with an attention-based architecture for pediatric respiratory disease classification. Scalogram images were generated from both full recordings and individual breath events to capture multi-resolution representations of respiratory sounds. To mitigate class imbalance, class-wise focal loss was applied during model training.\\
\textbf{Results:} The proposed model achieved an overall score of 0.9039 in binary event classification and 0.8448 in multiclass event classification. At the recording level, the model obtained scores of 0.720 for ternary classification and 0.571 for multiclass classification. These results outperform the previous best-performing models by 3.81\% and 5.94\%, respectively.\\
\textbf{Conclusion:} Our findings demonstrate that the proposed hybrid CNN–Transformer framework effectively captures the unique acoustic features of pediatric lung sounds. This approach offers a scalable and robust solution for automated pediatric respiratory disease diagnosis, with strong potential for deployment in resource-limited healthcare settings.

\end{abstract}






\section{Introduction}
The respiratory system plays a pivotal role in human health, making it crucial to accurately diagnose respiratory diseases (RDs) and monitor patient conditions. Common RDs include asthma, chronic obstructive pulmonary disease (COPD), pneumonia, and bronchitis, which are the third leading cause of death in the world\cite{ihme2023}. Early diagnosis of RDs is crucial for effective treatment and improving patient survival rates. In clinical medicine, lung auscultation has long served as a foundational method for the assessment of respiratory health. It is typically the first diagnostic step performed by healthcare professionals when evaluating patients with respiratory complaints due to its simplicity, non-invasiveness, and cost-effectiveness. Using a stethoscope, clinicians can detect and interpret a wide range of adventitious respiratory sounds (ARS), such as wheezes, crackles, rhonchi, and stridor, which often serve as early indicators of respiratory disorders~\cite{reichert2008analysis, bohadana2014fundamentals}. These sounds reflect underlying pathophysiological changes in airflow, secretion dynamics, or airway obstruction and can guide timely clinical decisions even before more advanced diagnostic tools are employed.

Despite its convenience, traditional auscultation has significant limitations. The diagnostic outcome is highly dependent on the level of experience of the clinician, auditory sensitivity, and training. Furthermore, external noise, patient variability, and subtle acoustic differences often contribute to inter-observer inconsistency, making it challenging to standardize auscultation across practitioners and healthcare settings. In resource-limited environments where access to specialized care and advanced imaging is constrained, this variability further compounds diagnostic uncertainty~\cite{gurung2011computerized, babu2022multiclass}. To address these challenges, the integration of electronic stethoscopes with artificial intelligence (AI)-based analysis systems has garnered increasing attention. Electronic stethoscopes allow for high-fidelity, noise-reduced recording of lung sounds, and can store, transmit, and replay auscultation data. When paired with AI algorithms these systems can extract robust features from lung sound signals, enabling automatic detection and classification of pathological events with high accuracy and consistency~\cite{moberg2025lung}. Unlike traditional auscultation, AI-driven systems are not limited by human perceptual boundaries, and offer scalable, repeatable, and objective assessments that reduce the reliance on specialist interpretation. These technologies hold substantial promise for improving respiratory care, particularly in rural clinics, emergency diagnosis settings, and telehealth platforms where rapid decision making and limited expert availability are key constraints. In addition, AI-augmented auscultation systems can serve as helper tools for medical training and second opinion generation, supporting clinical decision-making even in well-funded environments. The convergence of portable auscultation hardware and intelligent diagnostic algorithms thus marks a significant step toward democratizing respiratory health assessment and achieving more equitable healthcare outcomes globally.

\begin{table*}[!t]
    \centering
    \caption{Summary of the recent deep learning-based works on lung sound classification }\label{LSCX}
    \renewcommand{\arraystretch}{1.7}
    \begin{large}
        
\resizebox{\linewidth}{!}{
\begin{tabular}{p{3cm}|p{3cm}|p{4.5cm}|p{3.5cm}|p{4cm}}
\hline
\hline
\textbf{Study} & \textbf{DL Technique} & \textbf{Input Features} & \textbf{Disease Classification} & \textbf{Performance Metrics} \\
\hline
\hline

Dokur et al.~\cite{dokur2003classification} (2003)& MLP & MFCC, Spectrogram & COPD  & Precision: 0.90, Recall: 0.85 \\
\hline
Nishi et al.~\cite{haider2019respiratory} (2019) & SVM, KNN, DT  & Lung sound \& spirometry features & COPD   & Acc. 83.6\%\\
\hline
Acharya et al.~\cite{acharya2020deep} (2020)& CNN-LSTM & Mel-spectrograms & Breathing cycles  & ICBHI score 66.31\% \\
\hline
Shuvo et al.~\cite{shuvo2020lightweight} (2020) & Lightweight CNN & Scalograms & Pathological Condition & ICBHI score 98.70\%   \\
\hline
Basu et al.~\cite{basu2020respiratory} (2020) & GRU & MFCC & Pathological  & Acc.: 95.25\% \\
\hline
Nguyen et al.~\cite{nguyen2020lung} (2020) & CNN Snapshot Ensemble & Mel-spectrogram & Breathing cycles &\makecell[l]{Acc.: 78.4\% (4-class) \\ 83.7\% (2-class)} \\
\hline
Sudha et al.~\cite{jayalakshmy2021conditional} (2021)& Transfer learning &  Scalogram & Breathing cycles & \makecell[l]{Acc.: 92.50\% (ICBHI) \\ 92.68\% (RALE)} \\
\hline
Tariq et al.~\cite{tariq2022feature} (2022) & DCNN & Spectrogram, MFCC, Chromagram & Pathological Condition & Acc. 99.1\% \\
\hline
Brunese et al.~\cite{brunese2022neural} (2022)& ANN & Statistical Feature Set & Pathological Condition & Acc. 98.0\% \\
\hline
Semmad et al.~\cite{semmad2023scalable} (2023)& CNN & Mel-spectrogram & Respiratory disorders & Acc. 98.5\% \\
\hline
\hline
\end{tabular}}
\end{large}
\end{table*}

Currently, there are two main types of AI-based solutions for the classification and detection of lung anomalies from sound. Earlier approaches primarily relied on feature engineering methods, where handcrafted feature sets such as Mel-frequency cepstral coefficients (MFCCs)~\cite{ bahoura2003new}, wavelet transforms (WT)~\cite{ bahoura2009pattern}, and short-time Fourier transform (STFT) spectrograms~\cite{ acharya2017feature} were extracted and fed into machine learning classifiers. Traditional models such as Hidden Markov Models (HMMs)~\cite{ okubo2014classification}, Support Vector Machines (SVMs)~\cite{ gronnesby2017feature}, and Decision Trees~\cite{ chambres2018automatic} were widely used for classifying adventitious respiratory sounds, including wheezes, crackles, and rhonchi using these features. While these approaches achieved reasonable performance, their reliance on handcrafted feature extraction methods limited their adaptability.
A more recent group of studies has focused on transforming audio recordings into two-dimensional representations while employing deep learning techniques. These data-driven deep learning approaches eliminate the need for manual feature selection~\cite{pham2021cnn, ngo2021deep, minami2019automatic, perna2019deep,shuvo2020lightweight}. For instance, convolutional Neural Networks (CNNs)~\cite{ pham2021cnn, ngo2021deep} with spectrogram-based representations of respiratory sounds enable automated learning of relevant audio patterns. In contrast, recurring neural networks (RNNs) and attention mechanisms have been used to capture long-term temporal dependencies in respiratory signals, further improving the classification accuracy~\cite{ minami2019automatic, perna2019deep}. In ~\cite{ shuvo2020lightweight }, lightweight architectures based on hybrid scalograms have demonstrated promising results in detecting abnormal respiratory sound events. A short summary of deep learning-based work on respiratory sound classification is provided in Table~\ref{LSCX}.


The pediatric population is affected by a variety of respiratory conditions, with pediatric pneumonia remaining a particularly critical public health concern. It accounts for a substantial proportion of hospitalizations and mortality in children under five\cite{tazinya2018risk,lanata2008acute}. Despite the severity of these conditions, the development of automated respiratory sound classification systems has primarily focused on adult populations. A major limiting factor has been the lack of publicly available datasets representative of younger age groups. The recent release of the SPRSound dataset~\cite{zhang2022sprsound}, specifically curated for pediatric patients, represents a significant milestone in this domain. ResNet~\cite{ chen2022classify, li2022improving}, Inception~\cite{ngo2023deep, ngo2023inception}, DL+ML~\cite{taghibeyglou2023fusion} based, architectures have been utilized to process spectrogram-based representations of pediatric respiratory sounds, allowing for automated learning of relevant audio patterns. The combination of Inception and ResNet architectures proposed by Ngo et al.~\cite{ngo2023inception} captures features at multiple scales, while ResNet provides residual connections to ease the training of deeper networks. This hybrid architecture allows the model to learn richer features from spectrograms, making it particularly well-suited for classifying complex respiratory sounds. To further improve model performance, Ngo et al.~\cite{ngo2023deep} proposed incorporating multi-head attention mechanisms. This technique allows the model to focus on different spectral and temporal features across multiple spectrograms simultaneously. The attention mechanism helps the model to prioritize the most informative parts of the lung sounds, improving the accuracy of classification, especially for subtle anomalies in the sounds. Li et al.~\cite{li2022improving} introduced Focal Loss, which helps improve the recall of these minority classes, making the classification model more robust in identifying rare but clinically important respiratory events. A key challenge in this dataset is class imbalance, where normal respiratory sounds are overrepresented compared to adventitious sounds. This imbalance skews the predictions of the model, reducing the sensitivity to abnormal cases. Various strategies have been proposed to address this problem, including data enhancement techniques such as MixUp ~\cite{ hu2023supervised}, SMOTE technique~\cite{ zhang2022feature}. More recently, contrastive learning techniques have been introduced to enhance feature representation by clustering similar respiratory events~\cite{ moummad2023pretraining }.

\begin{figure*}[h]
    \centering
    \includegraphics[width=\linewidth]{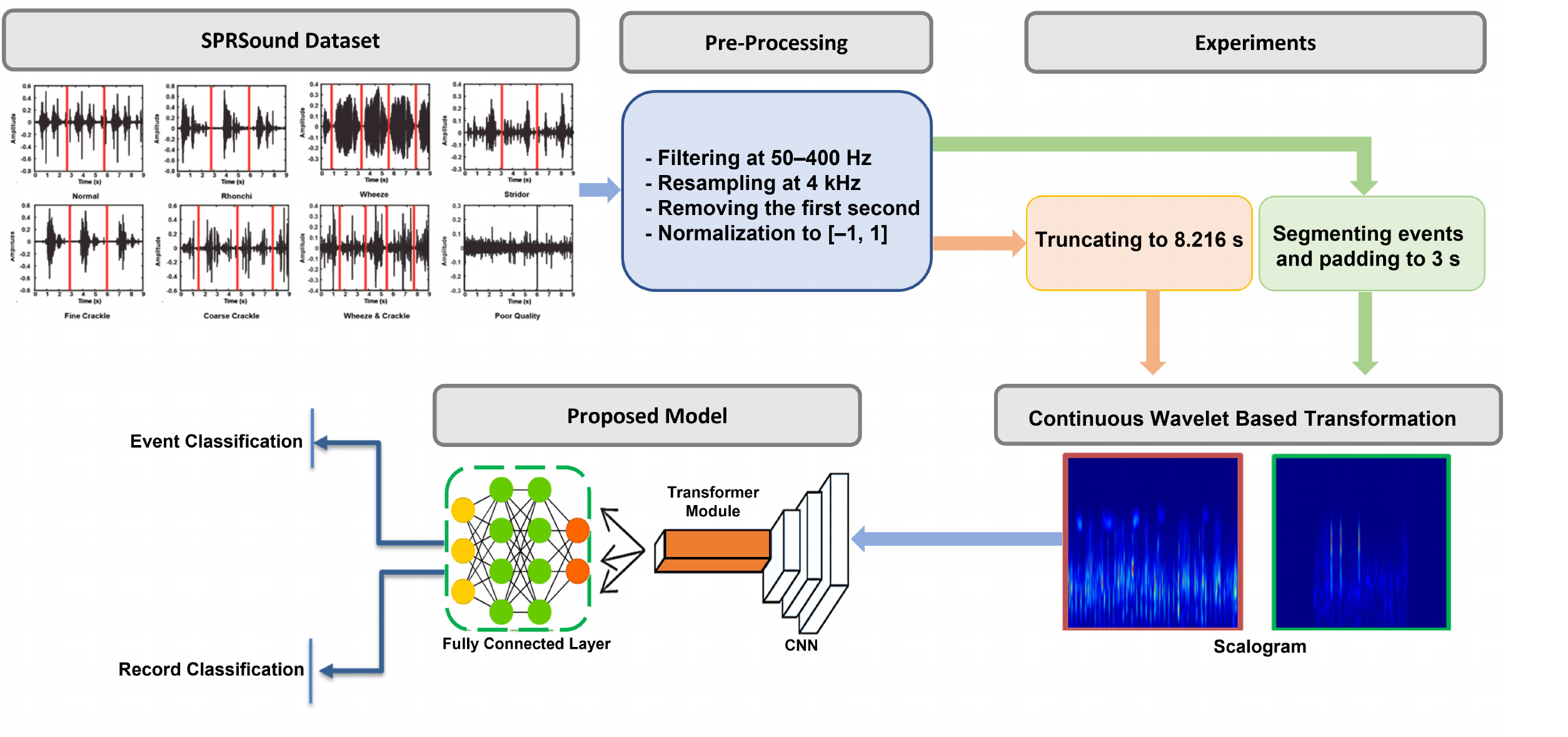}
    \caption{A graphical overview of the proposed framework for lung sound analysis. The framework begins with the SPRsound dataset, followed by several pre-processing steps, including low-pass filtering, resampling, and normalization. The lung sound signals are then segmented based on labels and padded to a fixed length. These transformed signals are then fed into the proposed model, which performs classification tasks}
    \label{GA}
\end{figure*}
\par  
Despite notable advancements in pediatric respiratory sound classification, most existing approaches predominantly concentrate on cycle-level classification, where each respiratory cycle is independently analyzed. While this granularity can aid in capturing local pathological patterns, it often overlooks the broader temporal dependencies and contextual information present across entire recordings. Consequently, such models struggle with real-world applicability, where clinical decisions typically rely on interpreting the overall respiratory pattern over extended periods, rather than isolated events. These limitations underscore the need for a more comprehensive framework that not only excels in identifying pathological events at a fine-grained level but also aggregates and contextualizes these events to make reliable recording-level decisions. Addressing these challenges, this work introduces a novel multi-stage hybrid model that couples efficient feature extraction with sophisticated feature-emphasizing and aggregation strategies to bridge the gap between event-level precision and recording-level reliability. A beirf overview of the study is illustrated in Figure~\ref{GA}.

\par 
The major contributions of this work are outlined as follows:
\begin{itemize}
\item Proposed a multi-stage hybrid model that integrates the efficiency of MobileNetV2 for feature extraction with the Transformer-based self-attention mechanisms for feature emphasizing, which improves the classification of pediatric respiratory sounds.
\item Incorporated a class-weighted sparse categorical focal loss, which gives lower weights to easily classified examples and focuses on harder-to-classify cases, mitigating the class imbalance present in respiratory sound datasets.
\item Conducted extensive experiments and compared the performance of the proposed model against existing state-of-the-art systems, demonstrating superior accuracy and robustness, especially in the classification of both cycle-level and recording-level tasks.
\item Conducted experiments to demonstrate how the feature emphasizing block enhances the model’s ability to differentiate between respiratory sound classes, improving interpretability and decision-making.

\end{itemize}
The rest of the paper is organized as follows. Section~\ref{mt3&4} provides an overview of the data resources, the pre-processing methods and the feature extraction method, followed by the description of the proposed novel architecture in Section~\ref{mt5}. Section~\ref{mt6} outlines the experimental setup, and the results are presented and discussed in Section~\ref{mt7}. Finally, the limitations and future scope of the work are presented in Section~\ref{mt8}, followed by the conclusion in Section~\ref{mt9}.

\section{Materials and Methods}\label{mt3&4}
\subsection {Dataset}
In this study, we utilized the open-access SPRSound pediatric respiratory sound database developed by Shanghai Jiao Tong University~\cite{zhang2022sprsound}. The dataset has been used for the IEEE BioCAS 2022 challenge. This database is the first open-access pediatric respiratory sound database, addressing a gap in existing datasets that predominantly focus on adult populations. This database consists of 2,683 records and 9,089 respiratory sound events from 292 participants~\cite{zhang2022sprsound}. The database categorizes the data at two levels: Record Level (represents the classification of the entire respiratory sound record) and Event Level (individual respiratory sound events within the recordings).
A detailed class distribution is shown in Table~\ref{tab:sprsound_dataset}.
\begingroup
\setlength{\tabcolsep}{8pt} 
\renewcommand{\arraystretch}{.8} 

\begin{table}[!h]
\centering
        \caption{SPRSound Dataset Class Distribution for Training and Testing}
\resizebox{\columnwidth}{!}{%

\begin{normalsize}

    \begin{tabular}{l|l|c|c}
        \hline
        \hline
        \textbf{Level} & \textbf{Type} & \textbf{Training} & \textbf{Testing} \\
        \hline
        \hline
        \multirow{5}{*}{Record} 
        & Normal & 1303 & 482 \\
        & Continuous adventitious sounds (CAS) & 126 & 107 \\
        & Discontinuous adventitious sounds (DAS) & 248 & 99 \\
        & CAS \& DAS & 95 & 36 \\
        & Poor Quality & 177 & 10 \\
        \hline
        \hline
        & \textbf{Total} & \textbf{1,949} & \textbf{734} \\
        \hline
        \hline
        \multirow{7}{*}{Event} 
        & Normal & 5159 & 1728 \\
        & Rhonchi & 39 & 14 \\
        & Wheeze & 452 & 413 \\
        & Stridor & 15 & 2 \\
        & Coarse Crackle & 49 & 17 \\
        & Fine Crackle & 912 & 255 \\
        & Wheeze \& Crackle & 30 & 4 \\
        \hline
        & \textbf{Total} & \textbf{6,656} & \textbf{2,433} \\
        \hline
        \hline
    \end{tabular}
          \end{normalsize}
   }

    \label{tab:sprsound_dataset}
\end{table}

\endgroup
\par

\subsection {Data Pre-processing}
\subsubsection{Bandpass filtering}
Lung auscultation signals typically fall within the frequency range of 50 to 2000 Hz~\cite{reichert2008}. To pre-process the audio samples of lung sound, a 4th-order Butterworth bandpass filter is applied, with the lower and upper cutoff frequencies set at 50 Hz and 2000 Hz, respectively. In order to maintain consistency and preserve key lung sound features while minimizing computational load, all audio samples are resampled at a frequency of 4 kHz. In addition, the signals are normalized in amplitude to mitigate the variations caused by different devices or sensors.

\subsubsection{Segmentation}
The dataset primarily contains audio samples of two durations: 9.216 seconds (1924 samples) and 15.216 seconds (759 samples). To pre-process the data, the first second of each recording is removed to eliminate the initial friction sound from the stethoscope. After this, the remaining 8.216-second segment is used for analysis. For the 15.216-second samples, we select either the first 8.216 seconds (0–8.216) or the last 8.216 seconds (7–15.216), depending on which segment contains more respiratory events. At the event level, each lung sound recording is segmented according to the timing of the annotated respiratory cycles. The cycle samples are then transformed into homogeneous signals with a fixed duration of 3 seconds with  zero padding~\cite{pham2020robust}.
\subsection{Transformation into 2D images}
The Continuous Wavelet Transform (CWT) is a signal processing technique that allows the decomposition of a signal into an orthonormal wavelet basis or a set of independent frequency components~\cite{gautam2013wavelet, debbal2004}. By using a basis function, i.e. the mother wavelet \( g(t) \), and it is scaled and shifted versions, the CWT can decompose a finite-energy signal \( x(t) \) as follows~\cite{meintjes2018}: 
\begin{equation}
Z(a, b) = \frac{1}{\sqrt{b}} \int x(t) g\left( \frac{t - a}{b} \right) dt
\end{equation}
Here, \( a \) and \( b \) represent the scale and translation parameters, respectively. The squared magnitude of the CWT coefficients \( Z \) is referred to as the scalogram~\cite{gautam2013wavelet}. Both the 8.216-second (recorded) and 3-second (segmented) lung sound samples are analyzed in the wavelet domain using the Morse analytic wavelet as the mother wavelet, and the scalogram images of \(224 \times 224\) resolution are produced. The time-bandwidth product and symmetry parameter for the wavelet are set to 60 and 3, respectively. The range of energy for the wavelet in both time and frequency determines the minimum and maximum scales, which are automatically chosen using 10 voices per octave~\cite{Continuo93:online}.

\begin{figure*}[t!]
    \centering
    \includegraphics[width=\textwidth]{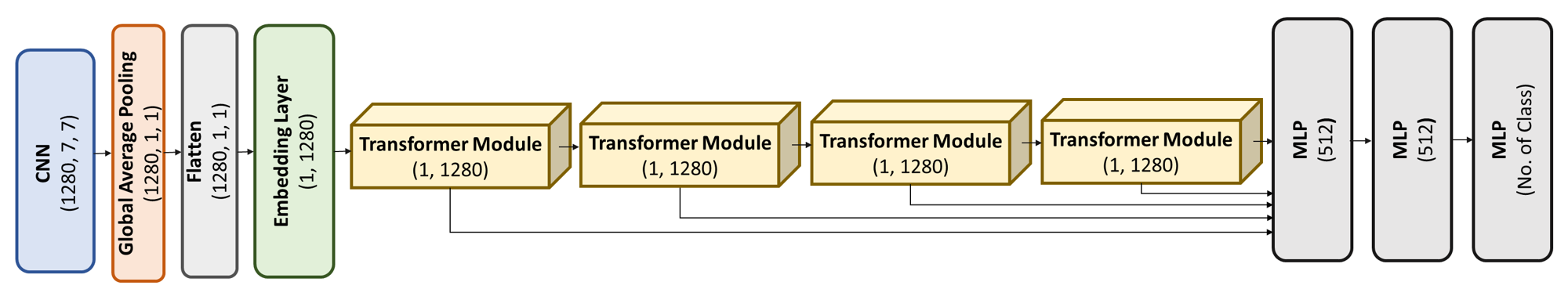}
    \caption{Architecture of the proposed  Multi-Stage Hybrid CNN-Transformer model for respiratory sound classification. The model consists of a MobileNet feature extractor, followed by global average pooling, an embedding layer, and a series of Transformer modules ( 8 attention heads with a hidden size of 2048). The final classification is performed by a multi-layer perceptron (MLP) to predict the diagnostic classes.}
    \label{f2}
\end{figure*} 
\section{Proposed Classifier Architecture}\label{mt5}

The proposed model is designed for the classification of lung sound scalograms. This architecture effectively combines a lightweight convolutional feature extraction network (with the network selection process described in Section~\ref{selection}), self-attention-based feature refinement, and a robust classifier to achieve efficient and accurate respiratory sound classification. The integration of depthwise separable convolutions and transformer-based attention enables the model to process spectro-temporal lung sound representations with improved generalization and computational efficiency. The overall architecture is illustrated in Figure~\ref{f2}.

\subsection{Feature Extraction Block}
The feature extraction block utilizes MobileNetV2~\cite{howard2017mobilenets}, a lightweight CNN optimized for efficient feature extraction. MobileNetV2 employs depthwise separable convolutions, significantly reducing computational complexity while retaining rich feature representations. The input to the model consists of 2D scalogram representations of lung sound signals with dimensions \( 224 \times 224 \times 3 \). The first convolutional layer applies a \( 3 \times 3 \) kernel with 32 filters, followed by a series of inverted residual blocks with linear bottlenecks to enhance feature expressiveness. The depthwise separable convolution is mathematically formulated as follows:
\begin{equation}
Y = (X \ast D) \ast P
\end{equation}
where:
\begin{itemize}
    \item \( X \) represents the input feature map,
    \item \( D \) is the depthwise convolution kernel,
    \item \( P \) is the pointwise convolution kernel,
    \item \( \ast \) denotes the convolution operation.
\end{itemize}

This operation significantly reduces the number of parameters compared to standard convolutions. The final output from MobileNetV2 is a feature map of size \( 7 \times 7 \times 1280 \), which is further refined in the next stage.

\subsection{Feature Emphasizing Block}
To enhance the extracted feature representations, the model incorporates a transformer-based self-attention mechanism. Unlike traditional CNNs, which rely on local spatial hierarchies, the transformer module captures global contextual dependencies across temporal and spectral dimensions~\cite{vaswani2017attention}. The feature maps from MobileNetV2 are first flattened and projected into a 1280-dimensional embedding space, which serves as the input to the multi-head self-attention module.

The self-attention mechanism is mathematically formulated as follows:
\begin{equation}
\text{Attention}(Q, K, V) = \text{softmax}\left( \frac{Q K^T}{\sqrt{d_k}} \right) V
\end{equation}
Where:
\begin{itemize}
    \item \( Q, K, V \) are the query, key, and value matrices,
    \item \( d_k \) is the dimensionality of the key vectors,
\end{itemize}

Our model employs eight attention heads, each computing an independent attention score, followed by a feedforward network with a hidden size of 2048, given by
\begin{equation}
\text{FFN}(x) = \max(0, x W_1 + b_1) W_2 + b_2
\end{equation}
where \( W_1, W_2 \) and \( b_1, b_2 \) are learnable parameters. Layer normalization is applied for stable training, and dropout is used to prevent overfitting. Four such transformer modules in series constitute the feature emphasizing block.

\subsection{Classifier}
The refined feature representations from the transformer modules are passed through the classifier, which consists of three fully connected layers. The first two fully connected layers have 512 and 256 neurons, respectively, sandwiched by a dropout layer with probability \( p = 0.3 \). The final classification layer consists of \( C \) output neurons corresponding to the number of target classes and applies the softmax activation function to compute class probabilities:
\begin{equation}
P(y_i) = \frac{e^{z_i}}{\sum_{j=1}^{C} e^{z_j}}
\end{equation}

where, \( z_i \) is the logit for class \( i \), \( C \) is the total number of classes and \( P(y_i) \) represents the probability of class \( i \).

The entire model is trained using the class-weighted sparse categorical focal loss function. 

\section{Experimental Setup}\label{mt6}
\subsection{Task Definitions}
The IEEE BioCAS 2022 challenge introduces two main tasks~\cite{zhang2022sprsound}. In this study, we opted for these tasks. Task 1, which focuses on the classification of sound events, is divided into two subtasks: Task 1-1 and Task 1-2. Task 1-1 is responsible for categorizing respiratory sound events as either Normal or Adventitious. Task 1-2, on the other hand, further classifies these events into more specific categories, including Normal, Rhonchi (Rho), Wheeze (W), Stridor (Str), Coarse Crackle (CC), Fine Crackle (FC), and Wheeze and Crackle (WC). Task 2, which addresses the classification of the entire respiratory recording, consists of Task 2-1 and Task 2-2. Task 2-1 focuses on classifying the recording as Normal, Adventitious , or Poor Quality (PQ). Task 2-2 is a multi-class classification task, where recordings are classified into categories such as Normal, CAS (Continuous Adventitious Sounds), DAS (Discontinuous Adventitious Sounds), CAS \& DAS, or PQ.
\subsection{Evaluation Criteria}
The training audio files (both for recording level and event level classification) are split into two sets with an approximate proportion of $90:10$ split for creating training and validation sets. The provided testing set defined by the challenge was used to evaluate the model. 
\par 
To align with the evaluation metrics of the IEEE BioCAS 2022 challenge, the tasks and sub-tasks in this study are evaluated based on the following metrics: sensitivity (SE), specificity (SP), average sensitivity (AS), harmonic sensitivity (HS), and the overall score. These metrics are calculated as follows:

\begin{equation}
    \text{SE} = \frac{T_i}{\sum N_i}~and~\text{SP} = \frac{T_N}{N_N} 
\end{equation}

where \(T_i\) denotes the number of correctly classified samples of class \(i\), \(\sum N_i\) is the total number of samples of target class \(i\). Similarly, \(T_N\) and \(N_N\) represent the number of correctly classified normal samples and the total number of standard samples, respectively.

Next, the average score (AS), harmonic score (HS), and the overall score (Score) are computed as follows:

\begin{equation}
    \text{AS} = \frac{\text{SE} + \text{SP}}{2} 
\end{equation}

\begin{equation}
  \text{HS} = \frac{2 \times \text{SE} \times \text{SP}}{\text{SE} + \text{SP}}  
\end{equation}

\begin{equation}
    \text{Score} = \frac{\text{AS} + \text{HS}}{2} 
\end{equation}


\subsection{Hyperparameters}
The proposed model is constructed using TensorFlow backend and trained using NVidia K80 GPUs. In this study, a batch size of $128$ has been selected for all the Tasks. Optimization is performed using the Adam optimizer with a learning rate of 0.001.
\subsection{Loss Function}
To enhance the detection of rare classes of lung sounds while solving the class imbalance problem, we employed a class-weighted sparse categorical focal loss function. This loss function incorporates a focusing parameter $\gamma$ that guides the function to focus on complex examples more than easily classified examples, thereby improving model robustness to hard-to-classify cases~\cite{lin2017focal}. 
The focal loss is defined as follows:

\begin{equation}
L(y, \hat{\mathbf{p}}) = - w_y (1 - \hat{p}_y)^{\gamma} \log(\hat{p}_y)
\end{equation}

Where:  

\begin{itemize}
    \item $y \in \{0, \ldots, K - 1\}$ represents the ground truth class label for $K$ classes,
    \item $\hat{\mathbf{p}} = [\hat{p}_0, \hat{p}_1, ..., \hat{p}_{K-1}]$ represents the predicted probability distribution over k classes,
    \item $\gamma$ is the \textbf{focusing parameter}, which down-weights easy examples and emphasizes harder cases,
    \item $w_y$ is the class-specific weight used to counteract the class imbalance.
\end{itemize}

\section{Results and Discussion}\label{mt7}
\subsection{Selection of the Feature Extractor Model}\label{selection}

We experiment with four state-of-the-art models (MobileNetV2~\cite{howard2017mobilenets}, Inception-V2~\cite{szegedy2015going}, ResNetV2~\cite{he2016deep}, and VGG16~\cite{simonyan2014very}) to determine which model works best as the feature extractor in our model. These models were tested across both cycle-level and recording-level classification (as shown in the Table~\ref{tab:classification_results}). 
\begingroup
\setlength{\tabcolsep}{4pt} 
\renewcommand{\arraystretch}{1} 
\begin{table}[t!]
\centering
\caption{Performance Comparison of Different Feature Extractor Models for Recording-Level and Cycle-Level Classification Tasks}

\resizebox{\columnwidth}{!}{%
\begin{normalsize}
\begin{tabular}{lccccc|ccccc}
\hline
\hline
\multicolumn{11}{c}{\textbf{Cycle level Classification}} \\
\hline
\hline
 & \multicolumn{5}{c|}{\textbf{Task 1-1}} & \multicolumn{5}{c}{\textbf{Task 1-2}} \\
\cline{2-11}
 & SE & SP & AS & HS & Score & SE & SP & AS & HS & Score \\
\hline
\hline

\textbf{MobileNetV2} & \textbf{0.85} & \textbf{0.89} & \textbf{0.86} & \textbf{0.86} & \textbf{0.86} & \textbf{0.75} & \textbf{0.88} & \textbf{0.82} & \textbf{0.81} & \textbf{0.815} \\
\textbf{Inception V2} & 0.83 & 0.87 & 0.85 & 0.84 & 0.845 & 0.73 & 0.86 & 0.78 & 0.77 & 0.775 \\
\textbf{ResNetV2} & 0.81 & 0.86 & 0.79 & 0.78 & 0.785 & 0.74 & 0.87 & 0.80 & 0.79 & 0.795 \\
\textbf{VGG 16} & 0.78 & 0.84 & 0.81 & 0.80 & 0.805 & 0.72 & 0.85 & 0.77 & 0.76 & 0.765 \\
\hline
\hline
\multicolumn{11}{c}{\textbf{Recording level Classification}} \\
\hline
\hline
 & \multicolumn{5}{c|}{\textbf{Task 2-1}} & \multicolumn{5}{c}{\textbf{Task 2-2}} \\
\cline{2-11}
 & SE & SP & AS & HS & Score & SE & SP & AS & HS & Score \\
\hline
\hline
\textbf{MobileNetV2} & \textbf{0.710} & 0.675 & \textbf{0.690} & \textbf{0.685} & \textbf{0.688} & 0.425 & 0.701 & 0.563 & 0.529 & 0.546 \\
\textbf{Inception V2} & 0.695 & \textbf{0.690} & 0.680 & 0.675 & 0.678 & 0.460 & 0.725 & 0.590 & 0.560 & 0.575 \\
\textbf{ResNetV2} & 0.685 & 0.680 & 0.670 & 0.660 & 0.665 & \textbf{0.480} & \textbf{0.740} & \textbf{0.605} & \textbf{0.570} & \textbf{0.590} \\
\textbf{VGG16} & 0.670 & 0.665 & 0.660 & 0.650 & 0.655 & 0.435 & 0.710 & 0.570 & 0.540 & 0.555 \\
\hline
\hline
\end{tabular}%
\end{normalsize}
}
\label{tab:classification_results}
\end{table}
\endgroup
\par In cycle-level classification, MobileNetV2 shows the highest scores in both tasks, achieving a score of 0.815 in Task 1-2, with high sensitivity and specificity values. Similarly, in the recording level performance, MobileNetV2 achieved the best overall balance in terms of sensitivity and specificity. In Task 2-1, it showed strong performance with a score of 0.675 for sensitivity and 0.710 for specificity. This indicates that MobileNetV2 is effective at detecting relevant features in the data while keeping errors to a minimum. Although MobileNetV2 had a slightly lower score of 0.546 compared to other models in Task 2-2, it still performed fairly well in comparison. This consistent performance across both tasks highlighted its robustness and versatility with relatively low computational cost. Therefore, we chose MobileNetV2 as the feature extractor for our model.

\subsection{Ablation Study}
To evaluate the impact of the focal loss parameter $\gamma$ on classification performance, we conducted an ablation study across different $\gamma$ values ranging from 2 to 5. This analysis helps isolate and understand the role of $\gamma$ in tuning the loss function’s focus on hard-to-classify examples and providing insights into its influence across tasks of varying granularity. The results, detailed in Table~\ref{tab:ab}, highlight the sensitivity of both cycle-level and recording-level classification tasks to this hyperparameter.
\begingroup
\setlength{\tabcolsep}{4pt}
\renewcommand{\arraystretch}{1} 

\begin{table}[t!]
\begin{normalsize}
\centering
\caption{Performance Comparison of Proposed Model at Different $\gamma$ Values for Cycle and Recording-Level Classification}

\resizebox{\columnwidth}{!}{%
\begin{tabular}{lccccc|ccccc}
\hline
\hline
\multicolumn{11}{c}{\textbf{Cycle-Level Classification}} \\
\hline
 & \multicolumn{5}{c|}{\textbf{Task 1-1}} & \multicolumn{5}{c}{\textbf{Task 1-2}} \\
\cline{2-11}
& SE & SP & AS & HS & Score &  SE & SP & AS & HS & Score \\
\hline
\hline
\textbf{$\gamma=2$} & 0.891 & 0.910 & 0.900 & 0.900 & 0.900  & 0.765 & 0.909 & 0.837 & 0.831 & 0.838 \\
\textbf{$\gamma=3$} & 0.895 & 0.905 & 0.900 & 0.900 & 0.900  & 0.753 & \textbf{0.931} & 0.842 & 0.833 & 0.838 \\
\textbf{$\gamma=4$} & 0.891 & \textbf{0.914} & \textbf{0.904} & \textbf{0.904} & \textbf{0.904}  & \textbf{0.783} & 0.912 & \textbf{0.847} & \textbf{0.842} & \textbf{0.845} \\
\textbf{$\gamma=5$} & \textbf{0.903} & 0.905 & 0.902 & 0.902 & 0.900 &  0.769 & 0.901 & 0.839 & 0.830 & 0.832 \\
\hline
\hline
\multicolumn{11}{c}{\textbf{Recording-Level Classification}} \\
\hline
 & \multicolumn{5}{c|}{\textbf{Task 2-1}} & \multicolumn{5}{c}{\textbf{Task 2-2}} \\
\cline{2-11}
 & SE & SP & AS & HS & Score & SE & SP & AS & HS & Score \\
\hline
\hline
\textbf{$\gamma=2$} & \textbf{0.726} & 0.670 & 0.697 & 0.696 & 0.696  & \textbf{0.446} & 0.716 & 0.581 & 0.550 & 0.566 \\
\textbf{$\gamma=3$} & 0.612 & \textbf{0.807} & 0.709 & 0.698 & 0.706 & 0.418 & \textbf{0.780} & \textbf{0.599} & \textbf{0.544} & \textbf{0.571} \\
\textbf{$\gamma=4$} & 0.665 & 0.718 & 0.692 & 0.691 & 0.691 & 0.397 & 0.757 & 0.577 & 0.526 & 0.549 \\
\textbf{$\gamma=5$} & 0.590 & 0.843 & \textbf{0.730} & \textbf{0.710} & \textbf{0.720}  & 0.434 & 0.734 & 0.584 & 0.546 & 0.565 \\
\hline
\hline

\end{tabular}%
}
\end{normalsize}
\label{tab:ab}
\end{table}

\endgroup
\par In the cycle-level classification task, our model demonstrated promising results in distinguishing between different respiratory sounds’ characteristics across both Task 1-1 and Task 1-2 (as seen in Table~\ref{tab:ab}). In Task 1-1, a $\gamma$ value of 4 yielded the best performance, achieving a score of 0.9039, which showed a 0.41\% improvement over the baseline $\gamma=2$ model, indicating its efficacy in capturing key features of the respiratory sounds. The improvement in performance was more noticeable in Task 1-2, where $\gamma=4$ again outperformed the other $\gamma$ configurations, achieving an overall score of 0.8448, which was a 0.85\% improvement over $\gamma=2$. Similarly, high sensitivity and specificity values reflect the model’s superior ability to generalize across various adventitious sound types. 
\par Due to the challenges of noise and variability across entire recordings, the proposed model’s results for the recording level classification task demonstrated a more varied performance across the different $\gamma$ values (as seen in Table~\ref{tab:ab}). In Task 2-1, $\gamma=5$ achieves the highest score of 0.720, with a significant specificity gain (SP = 0.843, highest among all models), indicating its strong ability to reduce false positives. However, in Task 2-2, $\gamma=3$ outperforms others with a score of 0.571, showing a 0.89\% improvement over the baseline $\gamma=2$ (0.566) configuration.

\par The performance gap between cycle-level and recording-level tasks is evident, with cycle-level accuracy surpassing recording-level accuracy by 9–15\%, highlighting the difficulty of handling extended respiratory segments. 
Here, we chose $\gamma=4$ and $\gamma=3$ as optimal values of our proposed model for cycle-level and recording-level classification, respectively,

\subsection{Comparison with Existing Methods}
In this study, our proposed model is compared with existing state-of-the-art methods in the classification of respiratory sounds. Table~\ref{tabSOTA} presents a summary of this comparative analysis, highlighting key performance metrics in different classification tasks at the cycle and record levels.
\begingroup
\setlength{\tabcolsep}{4pt} 
\renewcommand{\arraystretch}{1} 

\begin{table}[t!]
\begin{large}
\centering
\caption{Performance Comparison of Proposed Model with Different SOTA Models for Cycle and Recording-Level Classification}

\resizebox{\columnwidth}{!}{%
\begin{tabular}{lccccc|ccccc}
\hline
\hline
\multicolumn{11}{c}{\textbf{Cycle level Classification}} \\
\hline
\hline
 & \multicolumn{5}{c|}{\textbf{task 1-1}} & \multicolumn{5}{c}{\textbf{Task 1-2}} \\
\cline{2-11}
 & SE & SP & AS & HS & Score & SE & SP & AS & HS & Score \\
\hline
\hline
        \textbf{Ngo et al.~\cite{ngo2023deep}} & 0.810 & 0.910 & 0.860 & 0.860 & 0.860 & 0.670 & 0.920 & 0.790 & 0.780 & 0.790 \\
        \textbf{Pham et al.~\cite{ngo2023inception}} & 0.840 & 0.855 & 0.849 & 0.849 & 0.849 & 0.678 & 0.883 & 0.781 & 0.767 & 0.774 \\
        \textbf{Moummad et al.~\cite{moummad2023pretraining}} & 0.890 & 0.900 & 0.890 & 0.890 & 0.890 & 0.680 & \textbf{0.940} & 0.810 & 0.790 & 0.800 \\
        \textbf{Chen et al.~\cite{chen2022classify}} & 0.890 & 0.900 & 0.890 & 0.890 & 0.890 & 0.680 & \textbf{0.940} & 0.810 & 0.790 & 0.800 \\
        \textbf{Proposed Model} & \textbf{0.891} & \textbf{0.914} & \textbf{0.904} & \textbf{0.904} & \textbf{0.904} & \textbf{0.783} & 0.912 & \textbf{0.847} & \textbf{0.842} & \textbf{0.845} \\
\hline
\hline
\multicolumn{11}{c}{\textbf{Recording level Classification}} \\
\hline
\hline
 & \multicolumn{5}{c|}{\textbf{Task 2-1}} & \multicolumn{5}{c}{\textbf{Task 2-2}} \\
\cline{2-11}
 & SE & SP & AS & HS & Score & SE & SP & AS & HS & Score \\
\hline
\hline
\textbf{Ngo~\cite{ngo2023deep}} & 0.670 & 0.760 & 0.710 & 0.710 & 0.710 & 0.400 & 0.760 & 0.520 & 0.520 & 0.550 \\
        \textbf{Pham et al.~\cite{ngo2023inception}} & 0.704 & 0.789 & \textbf{0.747} & \textbf{0.744} & \textbf{0.745} & 0.361 & 0.801 & 0.581 & 0.498 & 0.539 \\
        \textbf{Moummad et al.~\cite{moummad2023pretraining}} & \textbf{0.770} & 0.660 & 0.720 & 0.710 & 0.715 & 0.230 & \textbf{0.860} & 0.540 & 0.360 & 0.450 \\
        \textbf{Chen et al.~\cite{chen2022classify}} & 0.770 & 0.660 & 0.720 & 0.710 & 0.710 & 0.230 & 0.860 & 0.540 & 0.360 & 0.450 \\
        \textbf{Proposed Model } & 0.612 & \textbf{0.807} & 0.709 & 0.698 & 0.706 & \textbf{0.418} & 0.780 & \textbf{0.599} & \textbf{0.544} & \textbf{0.571} \\
\hline
\hline
\end{tabular}%
}
\end{large}
\label{tabSOTA}
\end{table}

\endgroup

\par Chen et al.~\cite{chen2022classify} employed a fine-tuned ResNet18 architecture utilizing STFT spectrogram features, achieving notable scores, particularly on Tasks 1-1 and 1-2. However, their method exhibits limitations in generalization, as evidenced by relatively lower scores in Tasks 2-1 and 2-2 (0.710 and 0.450, respectively). Moummad et al.~\cite{moummad2023pretraining} explored supervised contrastive learning that incorporates demographic metadata, improving generalization and anomaly detection. Despite demonstrating effective performance across multiple datasets, their approach still struggles with computational inefficiency and complexity arising from additional metadata dependencies, resulting in moderate scores (0.715 in Task 2-1 and 0.450 in Task 2-2). In contrast, our proposed method consistently demonstrates superior performance across both cycle and recording-level classifications. It achieves exceptional robustness and generalization, particularly evidenced by a superior overall Score of 0.904 in Task 1-1 and 0.845 in Task 1-2 also 0.720 and 0.571 in Task 2-1 and 2-2, respectively, significantly surpassing Chen et al. and Moummad et al. Although methods like those proposed by~\cite{ngo2023deep} and~\cite{ngo2023inception} are effective in controlled environments, they struggle with noise and environmental artifacts in clinical recordings. In Task 2-2, their results struggle to maintain accuracy in the presence of background noise variations in respiratory sound patterns, leading to unreliable predictions in noisy settings. However, our proposed method consistently demonstrates a 3.81\% and 5.94\% performance gain, respectively.

\begin{figure}[t!]
    \centering
    \includegraphics[width=\linewidth]{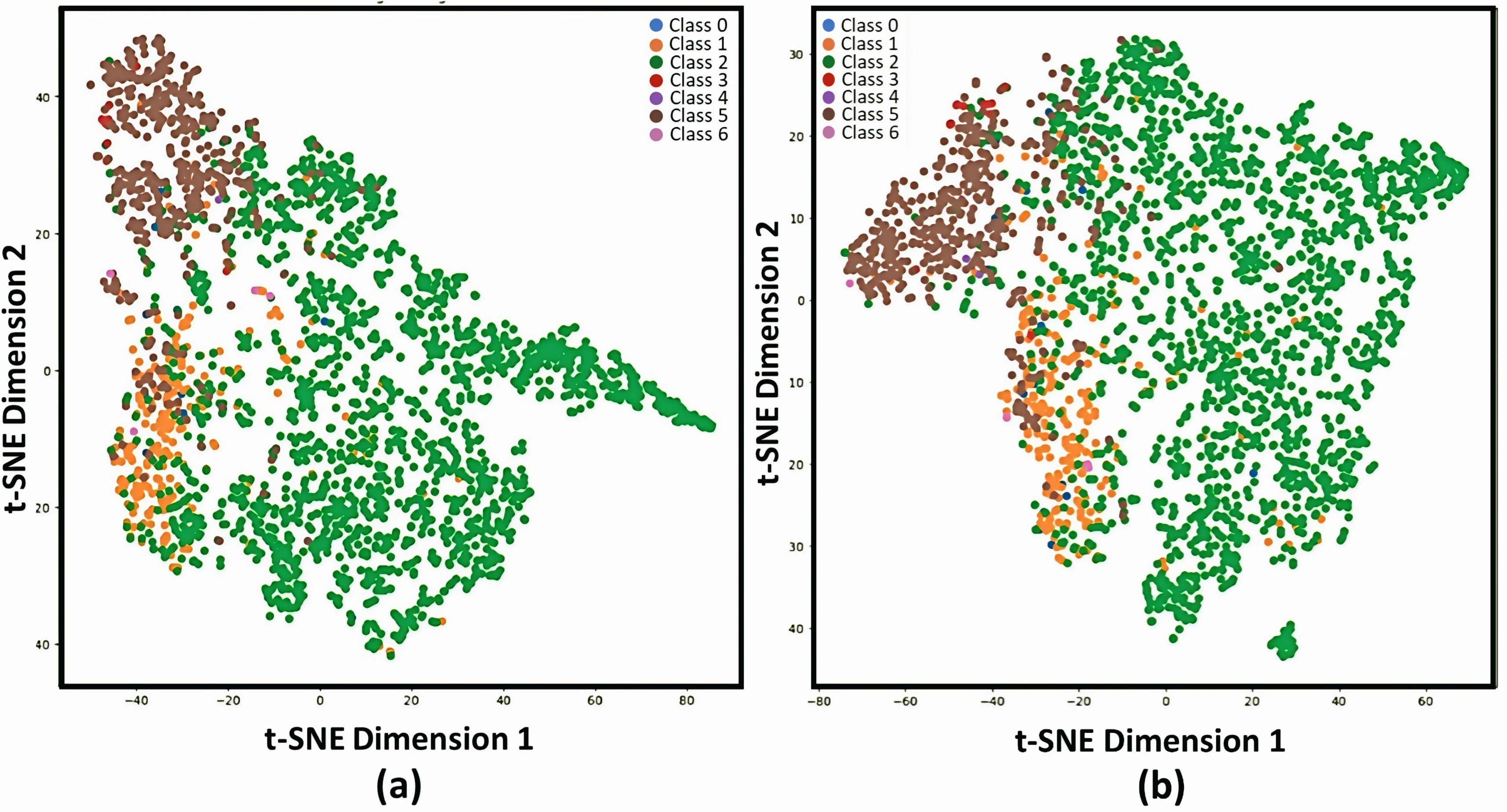}
    \caption{Two t-SNE plots of feature embeddings : one representing the MobileNetV2 model (left) and the other our proposed model (right) with incorporated feature emphasizing through transformer blocks }
    \label{tsne}
\end{figure} 
\subsection{Effect of Feature Enhancement on Embedding Space Separation}
In this study, we found that introducing a feature-enhancing block enhances the feature extraction process with MobileNetV2. To effectively analyze this, we compared two t-SNE (t-Distributed Stochastic Neighbor Embedding) plots of feature embeddings: one representing the MobileNetV2 model (left) and the other our proposed model (right) with incorporated feature emphasizing through transformer blocks (shown in Figure~\ref{tsne}). 

t-SNE plot works by mapping high-dimensional data points to a lower-dimensional space (typically 2D or 3D) while attempting to preserve the pairwise similarities between data points. Figure~\ref{tsne}~(a) demonstrated the t-SNE visualization for the MobileNetV2 model without feature emphasizing, highlighting its limitations in class separation. A significant overlap between multiple classes, particularly Class 0 (green) and Class 3 (red) and Class 4 (orange) and Class 5 (purple), have been observed in this plot. The clustering of these classes suggests that the model has difficulty learning discriminative features and fails to separate these classes in the higher-dimensional space. In contrast, the proposed model demonstrated a much more apparent separation between the classes (shown in Figure\ref{tsne}~(b)). Here, Class 0 (green) is now well-separated from Class 3 (red), and Class 2 (yellow) is clearly distinguished from Class 5 (purple). This enhanced separation in the t-SNE plot indicates that the feature-emphasizing mechanism helps the model to focus on the most discriminative features, improving the clarity of class boundaries and enhancing overall model performance in higher-dimensional space.

\section{Limitation and Future Work}\label{mt8}

While the proposed framework demonstrates promising results, its interpretability remains a significant concern for clinical deployment. The lack of transparency in decision-making processes can hinder clinicians’ trust and limit real-world integration. Enhancing explainability through interpretable machine learning techniques and visualization tools is essential to support clinical validation and user acceptance. Moreover, further optimization is necessary to enable real-time performance, especially in resource-constrained or point-of-care settings, where latency and computational efficiency are critical. Additionally, the current classification model has been trained on specific datasets, which may not comprehensively capture the diversity of respiratory conditions across different populations, age groups, and clinical scenarios. The limited availability of large-scale, well-annotated, and standardized datasets continues to be a major bottleneck in developing robust and generalizable models. 

Future research directions should explore the development of more sophisticated and context-aware frameworks that integrate a patient’s clinical history, presenting symptoms, and anatomical characteristics along with radiological image findings. Such holistic approaches can improve diagnostic accuracy and personalize treatment strategies. Moreover, there are scopes for analyzing other multimodal approaches using different types of physiological signals and imaging modalities.


\section{Conclusion} \label{mt9}

This study addressed the precise classification of pediatric respiratory diseases at both the event and recording levels. For that, we proposed a multi-stage hybrid CNN-Transformer model for classifying lung sounds, which effectively combines MobileNet-V2 for efficient feature extraction with transformer-based self-attention for enhanced feature refinement. The model demonstrated superior performance across event-level and recording-level tasks and has achieved a significantly high overall score of $0.9039$ and $0.8448$ for Task 1-1 and Task 1-2, respectively, in breath event classification. Similarly, the model outperformed all the SOTA systems at Task 2-2 with a score of $0.5710$. However, the model has limitations, including reliance on a relatively small dataset and high computational demands. 

\section*{CRediT authorship contribution statement}
\noindent \textbf{Samiul Based Shuvo:} Writing – review \& editing, Writing – original draft, Visualization, Validation, Software, Resources, Methodology, Formal analysis, Data curation, Conceptualization. 
\textbf{Taufiq Hasan:} Writing – review \& editing, Supervision, Project administration, Investigation.

\bibliographystyle{IEEEtran}
\bibliography{reference}\balance
\end{document}